\documentclass[12pt]{iopart}

\usepackage{graphicx}
\usepackage{amssymb}
\usepackage{ulem} 
\usepackage{verbatim}

\usepackage{color}
\usepackage{xcolor}
\usepackage{listings}
\usepackage{url}

\usepackage{caption}
\DeclareCaptionFont{white}{\color{white}}
\DeclareCaptionFormat{listing}{\colorbox[cmyk]{0.43, 0.35, 0.35,0.01}{\parbox{\textwidth}{\hspace{10pt}#1#2#3}}}

\graphicspath{{bilder/}}

\begin{document}

\title{Making the case of GPUs in courses on computational physics.}

\author{Knut S. Gjerden}
\address{Department of Physics,
Norwegian University of Science and Technology, N-7491 Trondheim,
Norway}

\ead{knut.skogstrand.gjerden@gmail.com}

\begin{abstract}
Most relatively modern desktop or even laptop computers contain a graphics card useful for more than showing colors on a screen. In this paper, we make a case for why you should learn enough about GPU (graphics processing unit) computing to use as an accelerator or even replacement to your CPU code. We include an example of our own as a case study to show what can be realistically expected.
\end{abstract}


\section{Introduction}
Using the graphics card in addition to, or in stead of the CPU to do high performance computing is a relatively new technology. Only the last few years has it been possible to run any algorithm through the parallel device previously specialized to handle only graphics to output on a screen. This paper argues for why physicists should begin to adapt this technology both in research applications and, as a natural extension, in teaching computational physics.

Modern physicists should have skills in theoretical analyses, experimental work and numerical modeling. Usually, one is mostly focused on one or two of these fields, but all three are important in physics. A theoretical physicist is different from a mathematician, an experimental physicist is different from an engineer, and a numerical physicist is different from a computational scientist. Being a physicist involves having a curious mind with the ability to put things into system and learn a broad range of, sometimes highly specialized, skills and techniques spanning multiple disciplines to accomplish this. As a consequence, the newest approaches and techniques in a field are not necessarily immediately adopted by the physics community, but have to reach a critical mass before they can be widely used. The purpose of this paper, is to convince you that performing numerical work using the graphics card is a technology that is  sufficiently mature that it is time for physicists to begin to use this approach. By extension, it is high time for this to be included in teaching so that the next generation of physicists can really take advantage of this form of parallelization. The future of computation is not serial, it is highly parallel.

As part of our case study, we port the discrete Fourier transform along with several vector functions to run on the graphics card. With porting the Fourier transform alone, we see about an order of magnitude in speedup in our application.

\subsection{What is a (GP)GPU}
Graphics processing unit (GPU), or general purpose GPU (GPGPU), computing is a term/phrase many perhaps have heard of but have limited experience of. The two terms are often mixed, but to little importance in our field. The graphics processing unit is a special unit in your computer traditionally tasked only to display graphics for the central processing unit (CPU), usually only to facilitate the use of a computer. The general purpose GPU is a special type of GPU which contains a functionally complete set of operations. If these operations can be performed on arbitrary bits in the GPU memory, the GPU can perform calculations as a CPU and is called a GPGPU. Thus, for computational work it is often assumed to be referring to a GPGPU whenever the term GPU appears.

Traditionally, the GPU is mounted on a separate board than the motherboard usually called a graphics card or video card. The two largest discrete manufacturers of graphics cards today are AMD and Nvidia. To give commands to the GPU, a programming language is required. The two most widely adapted languages are Nvidia's proprietary language CUDA, essentially an extension to C, and OpenCL, the Open Computing Language.

A GPU itself is a stream processor, designed to perform the same operation on a lot of similar elements. The basic design is a central control or managing unit which directs several compute units. A function on the GPU is called a kernel and it acts on all the elements in a stream. A typical kernel in, e.g., an unspecified matrix problem would look like in Listing \ref{kernelex}. These commands would be sent out to all compute units on the GPU which in turn would perform them on all elements streamed to them. From the contents of Listing \ref{kernelex}, we see that the ideal GPU data set is large, so that the first line takes as relatively short time as possible. It should also exhibit minimal dependencies on other parts of the set, allowing the second line to solve most of the problem without needing an advanced control structure of kernels. Additionally, the data set should have a high degree of parallelism allowing a greater number of elements in the stream.

\begin{lstlisting}[label=kernelex,caption=Kernel example - pseudocode]
 calculate what element in the matrix I am
 take my value and perform mathematics 
 return value
\end{lstlisting}

An analogy from baking: If you want to make flour, you don't want to crush single grains by hand, even if that gives you great control in how to crush each one; you want a machine that can take in a lot of similar grains at the same time and perform the same operation on all of them. Do not be mistaken, the GPU can handle very sophisticated problems, but it will perform better the better you are at formulating your problem as a parallel problem. Thinking parallel is different from thinking serial and looks to be an important skill as computers are becoming increasingly parallel. A skill that should be taught.

\subsection{And why should you care}
Interest in this type of computing has increased greatly over the last few years, to the point now where it is becoming common for manufacturers to put computing capable graphics cards into consumer products. A search on the websites of leading computer manufacturers and retailers at the time of writing among bestseller and basic consumer models, not gaming rigs or professional setups, yielded the odds greatly in favor of the graphics card being a GPGPU if a separate graphics was used. Almost all of the lowest end models used integrated graphics, a cheaper solution where the GPU is on-board together with the CPU and sharing RAM memory. Traditionally, this shared memory slows down the GPU to the point of not running well with a lot of computations, so it has not been a priority for the developers to support general purpose computing on these systems. To the authors best knowledge, only the newest generation of Intel processors support integrated graphics capable of this through OpenCL \cite{intel}.

The current state of affairs is that use of GPGPUs is on the rise and it might be a good idea for courses on computational physics to reflect this. Some may question the usefulness of GPGPUs and claim that this is a passing fad. Let us look at some basic facts. Games, both sports and recreational, have been a part of human history for over 5000 years. Versions of the egyptian game Senet has been found and dated to about 3100 BC \cite{senet}. Fast-forward till today, and the video game industry is a huge industry that has had a massively parallel problem as one of its core challenges for the last decades: Calculating the correct color of a pixel, based on increasingly complex algorithms, for millions of pixels more than thirty times per second. Advances in calculating skin color due to light reflecting in multiple skin layers for added realism and immersion in a video game is responsible for the boost in computing power computational scientists can harvest today. As long as there is a desire for increased realism in digital media, there will be advances in efficient parallel computation. The exact type of platform is impossible to predict, CPU-like, GPU-like or some new, possibly hybrid technology, but parallel computing is here to stay, and not constricted to massive supercomputers.

\subsection{The purpose of this paper}
GPGPU usage as an alternative to use of the CPU is often overlooked due to such feelings that it takes time to implement correctly, bug checking is more difficult, and the code becomes very platform dependent. Ideas such as these create an over dimensioned barrier of entry to use this technology. Let's get some of this out of the way instantly. If you cannot use a library and have to develop parallel code yourself, it usually does take time to implement correctly and bug checking is more difficult, irregardless of what processing unit you target. And to fully exploit the GPU available to you, you may have to write what could become platform dependent code. However, among the top five of the list over the top five hundred supercomputers in the world as of November 2011, three of them used GPUs \cite{top500}, so this is not necessarily a problem.

But you do not need a supercomputer to use GPUs, this is the main message of this paper. The three year old laptop I am writing this on has a dual core processor and a 48 core graphics card. A newer version laptop of comparable price point would use a quad core CPU, but could include a graphics card with over three hundred cores. It is not the purpose of this paper to advertise a purchase of a new graphics card, but rather to use what already may be available to you. The case study presented here will demonstrate how little effort it can take to turn a laptop or desktop computer into the computational equivalent of a cluster of computers.


\section{Problem definition}
\begin{figure}[htbp]
\begin{center}
\includegraphics[height=5cm]{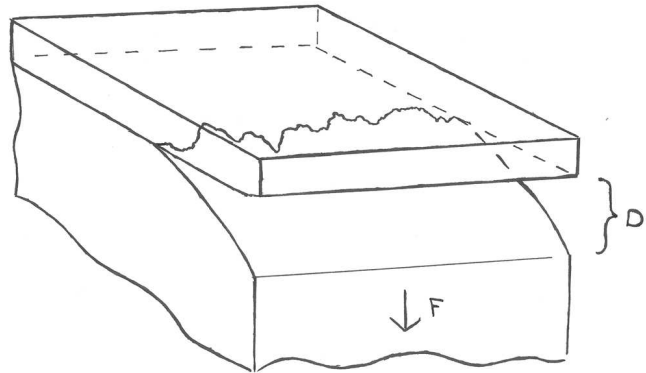}
\caption{Sketch of the model showing an elastic bulk material being initially connected to a stiff plate and pulled apart by a force $F$ at a global displacement $D$. The loading of such a system is usually executed by controlling either $F$ or $D$. A fracture front has developed at the interface between wholly connected and wholly separated material.}
\label{fig1}
\end{center}
\end{figure}

Let us set up our case study. We study the morphology of fracture fronts to learn more about fracture processes. The specific model is explained in detail in \cite{min, arne, min2}, but I will present the essence here. We model individual bonds connecting an elastic bulk material to a stiff plate or a mirror bulk system. The bonds behave linearly elastic, and forces acting on a specific bond is elastically transmitted through the material the bonds are connected to to all other bonds. The system is loaded through the stiff plate and we calculate the response in the elastic material. Conceptually, the model is illustrated in Figure \ref{fig1}. As can be inferred from the figure, the system can be either load-controlled or displacement-controlled. We choose the latter for simplicity. As will be shown, the natural derivations of the equations for this model leads to a matrix problem of type $\mathbf{A x} = \mathbf{b}$ where $\mathbf{x}$ is the load on the system and $\mathbf{A}$ contains the elastic response of the system resulting in the displacement $\mathbf{b}$. Solving this system for $\mathbf{x}$, rather than $\mathbf{b}$, entails using the matrix $\mathbf{A}$ and not its inverse.

Mathematically, the model is represented as an $L^2\times L^2$ matrix problem where the equations that govern each fiber are 
\numparts
\begin{eqnarray}\label{green}
  f_i=-k_i(u_i-D), \label{1a}\\
  u_i= \sum_jG_{ij}f_i, \label{1b}\\
  G^{}_{ij}= \frac{1-\nu^2}{\pi E a^2}\int\!\!\!\int^{a/2}_{-a/2}\frac{dx'dy'}{|\vec{r}_i(x,y)-\vec{r}_j(x',y')|}. \label{1c}
\end{eqnarray}
\endnumparts
Each fiber is such associated with a force it carries, $f_i$, a displacement of its rooting to the elastic material, $u_i$, due to the responses from all the other fibers, and an elastic constant, $k_i$. For simplicity, all fibers have the same elastic constant. In addition, each fiber is assigned a threshold value according to some random distribution which is used to determine the breaking point when the fiber is stretched. The shape of the elastic surface is given by (\ref{1b}), where the Green's function enters. This function relates how effects at one fiber $i$ affects another fiber $j$ at a distance $\vec{r}_i-\vec{r}_j$. $a$ is the lattice constant. In matrix notation, (\ref{1a}) and (\ref{1b}) can be re-written as
\begin{equation}\label{mateq}
(\mathbb{I} +\mathbb{KG})\vec{f}=\mathbb{K}\vec{D},
\end{equation}
where all vectors are of length $L^2$ and all matrices are of size $L^2\times L^2$.

For each step in the simulation, the displacement is set, and the resulting distribution of forces is calculated. For large enough local strain, bonds break and form damage which eventually develops to a fracture going through the material.
To get a directed fracture front in the system, we implement a gradient in the loading. This can be done in numerous ways. The numerically simplest one is to insert a gradient in the threshold distribution. Physically, this equates to controlling the angle of contact between the elastic material and the plate. 

\section{Numerical solution}
The primary purpose of the model just described was to study the morphology of the fracture front. As such there is considerable interest in studying as large systems as possible to get good statistics on the fracture front, which typically is of length longer than $L$, but much shorter than $L^2$. A naive solution of (\ref{mateq}) would scale in time as $L^4$, a very constricting behaviour that needs to be overcome.

The first thing to do is to exploit symmetries in the problem. The Green's function (\ref{1c}) is in essence just a two-point correlation function, and as such, only dependent on the distance between two such points. This has some major impacts on the problem. First, the coupling matrix $\mathbb{G}$ becomes symmetric and the whole matrix problem (\ref{mateq}) is positive-definite. This allows us to use an iterative solver like the conjugate gradient (CG) method, and the symmetry of $\mathbb{G}$ means we only have to store $L^2$ of data instead of the full $L^2\times L^2$. The result is that the CG solver only implicitly knows the coupling matrix and we only store the result of the matrix multiplication. This, in turn, effectively ensures that our problem becomes processor-dependent instead of memory-dependent. This is a great starting point for GPU-usage. The details of the solver structure are presented in Listing \ref{solver}.

\begin{lstlisting}[label=solver,caption=Solver - pseudocode]
 set up initial values
 start loop:
 	calculate A*x 								(vector-matrix mult.)
	calculate search length 			(dot product)
	calculate new x 							(vector-vector addition)
	calculate residue 						(vector-vector addition)
	calculate residue length	  	(dot product)
	update calculation variables	(vector-vector addition)
	check convergence
\end{lstlisting}

The second major impact of the Greens function is that it is diagonal in Fourier space, effectively reducing the matrix-vector multiplication to a vector-vector multiplication when using Fourier acceleration. This of course adds two Fourier transforms per iteration of the solver, but it is still faster than the complete matrix multiplication. The complete matrix problem can than be expressed as 
\begin{equation}\label{compmateq}
(\mathbb{I}+\mathbb{KF}^{-1}\mathbb{F}^1\mathbb{G})
\mathbb{F}^{-1}\mathbb{F}^1\vec{f}=\mathbb{K}\vec{D},
\end{equation}
in stead of (\ref{mateq}).

You may have noticed that all the steps in Listing \ref{solver} involve quite general function calls; all of them can be found in numerical libraries such as Basic Linear Algebra Subprograms (BLAS) \cite{blas}, accessible trough most programming languages. In our case, we used in addition the discrete Fourier transform library Fastest Fourier Transform in the West (FFTW) \cite{fftw}. These are all the pieces required for an efficient CPU implementation of problems of the same type as ours.

\subsection{GPU approach and implementation}
Numerical GPU libraries are not as widespread as their CPU counterparts yet. However, Nvidia has made CUBLAS and CUFFT available through their CUDA language \cite{cuda}. 
The graphics cards readily available at the time were manufactured by Nvidia, so that option was explored first. CUBLAS and CUFFT are modeled on BLAS and FFTW, so the interface is almost identical. The theoretical approach is thus simple: Replace CPU library calls with GPU library calls. The practical approach turned out to be not much more complicated. As this was the first attempt at GPU programming, the move from CPU to GPU was done in two steps. The first step was to move just the Fourier transforms onto the graphics card. 
Details of this step is in Listing \ref{fft}. 

\begin{lstlisting}[label=fft, caption=FFT using GPU - pseudocode]
	;Allocate memory on the GPU
	gpu-mem = cudaMalloc data
	;Copy data to GPU memory
	cudaMemcpy gpu-mem data host-to-device
	;Plan FFT
	cufftPlan size-of-data
	;Execute FFT on the GPU
	cufftExec plan gpu-mem direction
	;Copy result to CPU memory
	cudaMemcpy data gpu-mem device-to-host
	;Clean up to release GPU resources
	cufftDestroy plan
	cudaFree gpu-mem
	
	;Optional: Normalize data 
	;Data can either be normalized back in CPU memory, or using a call to CUBLAS when the data is still in GPU memory: 
	cublasCscal size-of-data normalizing-factor gpu-mem
\end{lstlisting}

%

Already an appreciable speed-up was obtained. Very few lines of code in the application needed to be changed, but a significant, immediate result was obtained. From function call ``CPU: Here is data, do FFT and return data'' to ``GPU: Here is data, do FFT and return data'', no GPU architecture dependencies have been introduced (apart from the choice of GPU programming language support), but the total run time of the code in question decreased by close to an order of magnitude.

Step two was to move the entire solver to run on the graphics card. In theory, one just has to replace certain function calls, mostly library calls to BLAS, but in practice some additional functions were needed to manage data as now part of the data is only known to the CPU and part is only known by the GPU. The most computationally expensive part of our application is the calculation of ``A*x'' in the beginning of Listing \ref{solver}. This is essentially a convolution, and the elements of it are listed in Listing \ref{fftconv}. We use real data types as much as possible to reduce memory usage and computation time, but use complex numbers for the Fourier routine in order to do in-place transforms to minimize data copying and movement. Both the input and output of the convolution consist solely of real numbers, as they should.

\begin{lstlisting}[label=fftconv, caption=Convolution of real data using in-place FFT on the GPU - pseudocode]
	;Copy trial x (real numbers) into complex form for FFT
	cpyComplex x buffer
	;Fourier transforming x
	cufft2D buffer :forward t
	;Performing element-by-element multiplication in Fourier space    (the FFT of G is static and performed during initialization)
	elementMult G-FFT buffer
	;Buffer now contains the result and is inverse transformed
	cufft2D buffer :forward nil
	;Returning the values as real data
	cpyReal buffer result
\end{lstlisting}

At the end of Listing \ref{fftconv}, ``result'' contains the result of the multiplication $\mathbb{G}\vec{x}$. One element-by-element multiplication and a call to CUBLAS later, and it contains the trial guess of the entire left hand side of (\ref{compmateq}). One of the function calls in this code snippet which is not a library call to CUBLAS, CUFFT or a standard function in CUDA is elementMult. This function is detailed in Listing \ref{code}. Remember the general example of Listing \ref{kernelex}. Function or variable declarations starting with a \textunderscore\textunderscore label\textunderscore\textunderscore~ denotes code accessible to the GPU and not the CPU. The first line in the kernel declaration calculates which element in the vector the current compute unit in the GPU is working on. The second and third lines ensures we are within bounds of the data, and the third and final line tells the compute unit to do complex multiplication of the corresponding elements of the complex vectors $\mathbf{A}$ and $\mathbf{B}$. The remainder of the code snippet defines a function callable from standard C which will execute the GPU kernel. The second and third lines could be replaced by a conditional, but the way GPUs are designed they are more efficient with direct calculation than branching code. Thus, conditionals should be avoided when possible and work should be assigned to thread calculation rather than loop structures. The example code is trivially tiny, but the concept warrants consideration.

\begin{lstlisting}[label=code, caption=GPU kernel example of destructive complex single float element multiplication of A * B into B - CUDA code]
const int threadsPerBlock = 256;
// declare kernel
__global__ void elementMult_kernel(int N, const cuFloatComplex* A,                                 cuFloatComplex* B)
{
    int i = blockDim.x * blockIdx.x + threadIdx.x;
    i = max(0, i);
    i = min(i, N);
    B[i] = cuCmulf(A[i], B[i]);
}
void elementMult(int N, const cuFloatComplex* A, cuFloatComplex* B)
{
 // invoke kernel
    int blocksPerGrid = (N + threadsPerBlock - 1) / threadsPerBlock;
    elementMult_kernel<<<blocksPerGrid, threadsPerBlock>>>(N, A, B);
}
\end{lstlisting}

\section{Results}
\begin{figure}[htbp]
\begin{center}
\includegraphics[width=13cm]{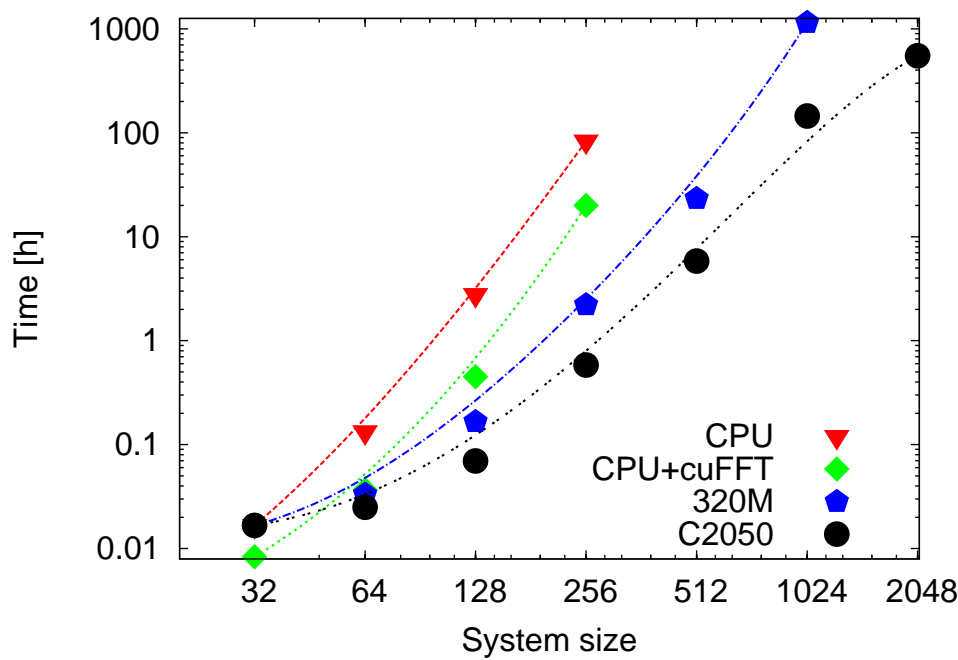}
\caption{Simulation wall time vs. system sizes plotted on a log-log scale. CPU denotes serial computer code only, cuFFT denotes Fourier transform performed on the 320M graphics card, and 320M and C2050 denotes full solver running on a graphics card of type 320M or C2050, respectively. Lines are guidelines to help illustrate the change in scaling behaviour. Further notes in text.}
\label{times}
\end{center}
\end{figure}

Only a handful of user-supplied functions were needed in addition to the numerical libraries readily available. The re-write to use the GPU was not complicated, especially if guidance is provided, and is a useful exercise to understanding GPU computation. From the CPU doing all the work, the situation now is that the graphics card does almost all the work, reporting back only a few values to the CPU each time the solver has completed its work. Simulation time for single systems are reported in Figure \ref{times}, where the effect of running on the GPU is clearly visible. Most of the timings were performed on on a Nvidia GeForce 320M in a laptop computer to illustrate the usually idle computing power available. Timings performed on a higher-end Nvidia Tesla C2050 are also included. The CPU used for the serial code was an 1.86GHz Intel Core 2 Duo.

Some quick notes regarding Figure \ref{times}. First, note that most of the data in the figure are based on simulations on a laptop computer, and all times are reduced when running on a more powerful computer. The times involving work on the graphics card would typically be reduced more than the serial CPU times as for processor intensive problems the clock frequency of the CPU is the limiting factor whilst on a more powerful graphics card the combination of clock frequency, number of cores and communication bandwidth determines the speedup. Usually, higher-end graphics cards have a greater number of cores and higher bandwidth, so even if the individual core clock frequency can be lower, the overall result is faster computation. Secondly, the data for 1024 and above are based on estimations, as no such systems have yet run to completion. A single system of this size would require just under seven weeks to complete, when GPU accelerated using the 320M. Thirdly, the curve for the full solver on the graphics cards starts up about equal to the pure CPU code, and then falls quickly below the other curves. The reason for this is that it takes some time to initialize the graphics card and move the initial data to GPU memory. 

Communication to and fro the CPU and GPU can quickly become a bottleneck as the internal bus on the GPU usually is an order of magnitude or more faster than communication between the GPU and the CPU. This became a key factor in deciding exactly what data should exist in the CPU memory and GPU memory. The next step up would be to put all the data on the GPU, but at that point it would probably be simpler to rewrite the entire code in CUDA or geared toward massive multicore CPU machines. There would be an additional speedup, of course, but a total rewrite of the code base was not desirable at this time.

Figure \ref{times} displays some very striking features of the GPU implementation. Comparing the curves for full solver and CPU only, we see that using the GPU, we can double the system size at no additional cost (forgetting about initialization time for the smallest systems). Remember that solving the original matrix problem in a straight-forward manner would scale in time as $L^4$. Perhaps most strikingly, looking at the data for $L=256$, the largest system size where we have data for all versions of the code, we see a speedup of two orders of magnitude if one takes advantage of the graphics card. It is also a point to be made that the scaling with respect to time is better for the GPU implementation than for only using the CPU, at least for $L<1024$, but remember that the time for 1024 is an estimate, not an actual timing. This is especially visible on the C2050. Presumably, the larger systems benefit from increased occupancy on the GPU with respect to my kernels, and decreased occupancy with respect to the library calls. What I mean is that the less efficient kernels written by me use more computation time for larger $L$, thereby masking the slower calls to memory as these can be performed asynchronously. Additionally, the library calls are more efficient at reusing data within a kernel, thus for larger $L$, these functions are better at using each thread for more work thus computing more for fewer calls to memory.

The overall result of this could in some cases mean the difference between running test systems and the capability of full-scale simulations on a desktop computer. This could have major implications for interactive teaching in lecture halls, classrooms or computer rooms. Imagine not being limited to demonstrating textbook models, but able to show relevant, state-of-the-art code to your students, possibly even running in real-time.

The attentive reader will probably have noticed that we use the conjugate gradient method without a pre-conditioner. If you are wondering why, the answer to that is that we used one for a while, but when we implemented some features described in \cite{min}, the pre-conditioner broke. We have not found a good replacement yet, but in total, less time was spent on transitioning the solver to run on the graphics card than finding a good pre-conditioner, and the speed-up achieved using the GPU is an order of magnitude greater than ever achieved through the old pre-conditioner.

\section{Conclusion}
The recommendation is clear:  Spare some time to examine if your problem have some elements that could be run on the graphics card and if you have a graphics card that can be used for general computing. Presenting some final numbers, the conclusion speaks for itself: The time used to implement code for the GPU was one to two weeks and a factor 100 in speedup was obtained. The CPU version of the code is estimated to solve a system of $L=512$ in just shy of three and a half months. The code using the GPU solves this system in under 24 hours.

\section*{Acknowledgements}
The author would like to acknowledge economic support from the Norwegian Research Council (NFR) through grant number 177591/V30. Part of this work was carried out under the HPC-EUROPA2 project (project number: 228398) with the support of the European Commission Capacities Area - Research Infrastructures Initiative. Part of this work made use of the facilities of HECToR, the UKÕs national high performance computing service, which is provided by UoE HPCx Ltd at the University of Edinburgh, Cray Inc and NAG Ltd, and funded by the Office of Science and Technology through EPSRCÕs High End Computing Programme.

\end{document}